\documentclass[aps,floatfix,twocolumn,a4paper,showpacs, nofootinbib,superscriptaddress,10pt]{revtex4}
\usepackage{graphicx,float}\usepackage{graphicx,float}
\usepackage[all]{xy}
\usepackage{amsmath,upgreek}
\usepackage{amssymb}
\usepackage{color}
\usepackage{epsfig,bm}		
\usepackage{graphicx,epstopdf}
\usepackage{subfigure}
\usepackage{pdfpages}
\usepackage[colorlinks,hyperindex]{hyperref}
 
  \newcommand{\clt}{\textcolor{black}}
\definecolor{green1}{RGB}{0,128,0} 
\hypersetup{backref=true,pagebackref=true}
\hypersetup{%
  colorlinks = true,
  linkcolor  = blue,
  citecolor = cyan,
}
\usepackage{bookmark,textgreek}
\usepackage{hyperref,color,xcolor}
\hypersetup{hidelinks,hyperindex=true,colorlinks=true,breaklinks=true,urlcolor= blue}
\hypersetup{%
  colorlinks = true,
  linkcolor  = blue
}\usepackage{amssymb}
\newsavebox{\foobox}
\newcommand{\slantbox}[2][0]{\mbox{%
    \sbox{\foobox}{#2}%
    \hskip\wd\foobox
    \pdfsave
    \pdfsetmatrix{1 0 #1 1}%
    \llap{\usebox{\foobox}}%
    \pdfrestore
}}
\newcommand\unslant[2][-.25]{\slantbox[#1]{$#2$}}

\newcommand{\upPartial}{\unslant\partial\kern-0.8pt}

\usepackage{graphicx,float,tikz}
\usepackage[all]{xy}
\newcommand\ringring[1]{%
  {% make an Ord atom
   \mathop{\kern0pt #1}\limits^{% set a box over the variable
     \vbox to-1.85ex{
       \kern-2ex % lower the ring accents
       \hbox to 0pt{\hss\normalfont\kern.1em \r{}\kern-.45em \r{}\hss}%
       \vss % fill
     }% end of \vbox
   }% end of the superscript
  }% end of \mathop
}\newcommand\orcidroldao{{\href{https://orcid.org/0000-0003-3978-532X}{\orcidicon}}}
\newcommand{\orcidicon}{%
	\begin{tikzpicture}
	\draw[lime, fill=lime] (0,0)
		circle [radius=0.16]
		node[white] {{\fontfamily{qag}\selectfont \tiny ID}};
	\draw[white, fill=white] (-0.0625,0.095)
		circle [radius=0.007];
	\end{tikzpicture}	\hspace{-2mm}
}
\newcommand{\bpartial}{\mathop{\partial\kern -4pt\raisebox{.8pt}{$|$}}}

\newcommand{\bes}{\begin{subequations}}
\newcommand{\ees}{\end{subequations}}
\def\beq{\begin{eqnarray}}
 
\def\eeq{\end{eqnarray}}
\def\be{\begin{equation}}
\def\ee{\end{equation}}

\begin{document}

\title{Information entropy in AdS/QCD: mass spectroscopy of isovector mesons}
\author{R. da Rocha\orcidroldao\!\!}
\email{roldao.rocha@ufabc.edu.br}
\affiliation{Federal University of ABC, Center of Mathematics, Santo Andr\'e, 09580-210, Brazil}

\begin{abstract}
The mass spectra of isovector $\Upsilon$, $\psi$, $\phi$, and $\omega$ meson resonances are investigated, in the AdS/QCD  and information entropy setups. The differential configurational entropy is employed to obtain the mass spectra of radial $S$-wave resonances, with higher excitation levels, in each one of these meson families, whose respective first undisclosed states are discussed and matched up to  candidates in the Particle Data Group.

 \end{abstract}
\maketitle

\section{Introduction}

In Shannon's theory, the information entropy of some random variable consists of the average level of information that is inherent to the possible outcomes of this variable \cite{Shannon:1948zz}. It can be performed by the theory  configurational entropy (CE), which comprises the amount of entropy endowing a discrete system that governs correlations among its constituents.  The CE  estimates how information can be encrypted in wave modes, in momentum space, still describing the whole system under scrutiny. The maximal CE is attained by a source having a  probability distribution that is uniform and, in this way, uncertainty attains the maximal value in the case of equiprobable events. 
When one takes the continuous limit, the differential configurational entropy (DCE) plays the role of the CE, once the information dimension\footnote{The information dimension  measures the  dimension of a probability distribution and evaluates the growth rate of the information entropy given by successively finer discretizations of the space. It also corresponds to the R\'enyi information dimension, as a  fundamental limit of almost lossless data compression for analog sources.} 
vanishes \cite{Bernardini:2016hvx}. 
The DCE has been employed to study physical systems in a variety of circumstances \cite{Gleiser:2018kbq,Gleiser:2011di,Gleiser:2012tu}, once a localized scalar field that portrays the system is taken into account \cite{Karapetyan:2018oye,Karapetyan:2018yhm}. 

In the precise context of AdS/QCD, the DCE has been playing a prominent role in the investigation of hadrons, estimating new features of hadronic  states  in AdS/QCD \cite{Bernardini:2018uuy,Ferreira:2019inu,daRocha:2021imz,Ferreira:2019nkz} and also corroborating to existing hadronic properties. Recent studies in the DCE setup of AdS/QCD investigated light-flavor mesons, heavy mesons encompassing charmonia and bottomonia resonances, tensorial mesons, baryons, glueball fields, pomerons, odderons, and baryons, including finite-temperature regimes  \cite{Colangelo:2018mrt,Ferreira:2020iry,Braga:2018fyc,Braga:2020myi,Braga:2020hhs,MarinhoRodrigues:2020yzh,Bernardini:2016qit,Braga:2017fsb,Karapetyan:2016fai,Karapetyan:2017edu,Karapetyan:2019ran,Karapetyan:2019fst,Karapetyan:2020yhs,Barbosa-Cendejas:2018mng,Ma:2018wtw}. Some of the obtained results were confirmed by the Particle Data Group (PDG) \cite{pdg} and also matched up to new meson candidates that were previously unattached to some of the studied meson families. Other relevant implementations of  the DCE paradigm in QFT and AdS/CFT  were reported in Refs.  \cite{Correa:2015vka,Correa:2016pgr,Cruz:2019kwh,Lee:2019tod,Bazeia:2018uyg,Bazeia:2021stz,Alves:2014ksa,Alves:2017ljt,Gleiser:2018jpd,Braga:2019jqg,Braga:2020opg,Casadio:2016aum,Fernandes-Silva:2019fez,Bernardini:2019stn,Lee:2018zmp,Thakur:2020sko}.

With the aid of data from experiments and theoretical developments, meson spectroscopy composes a paradigm asserting that mesons present ubiquitous features. Important studies support universal properties that include all the meson spectrum, running from light-flavor ones to heavy charmonium and bottomonium \cite{amsler,Simonov:1989fd,Badalian:2016ttl,Frederico:2014bva,dePaula:2009za}. Meson spectroscopy has been successfully scrutinized in the context of the AdS/QCD duality. In this setup, there is an AdS  bulk that places a weakly-coupled gravity sector, which is the dual theory of  conformal field theory (CFT) living on the codimension one AdS boundary. In the duality, physical fields that reside in the bulk are equivalent to operators in strongly-coupled QCD on the AdS boundary. 
The bulk extra dimension implements the QCD energy scale  \cite{Csaki,Karch:2006pv,Branz:2010ub}. Among possible models endowing AdS/QCD, the soft wall prescribes Regge trajectories to come out from confinement, being implemented by a dilaton that is a quadratic function of the energy scale  \cite{Colangelo:2008us,MartinContreras:2020cyg}. 
Linear Regge trajectories successfully describe the mass spectrum of light-flavor meson families, at least when considering the radial $S$-wave resonances that have not so high excitation $n$ levels. The range wherein the linearity holds can vary for each light-flavor meson family. When hadronic states are constituted by strange, charm, bottom, or top quarks, the linear regime $m_n^2\propto n$  is no longer verified \cite{MartinContreras:2020cyg,Afonin:2014nya}. 
The Bohr--Sommerfeld quantization
technique applied to the Salpeter-like 
equation was used to derive nonlinear Regge trajectories for meson states constituted by distinct quarks, presenting 
similar profiles as the ones for quarkonia \cite{MartinContreras:2020cyg},  complying with experimental data and also to the theoretical predictions \cite{Chen:2018nnr}.  Besides, due to the chasm among the heavy and light quark masses, more recent models regard massive quarks that constitute mesons. Therefore, more realistic scenarios can encompass nonlinear Regge trajectories, induced by the quark constituent mass. Light-flavor quarks comprehend up, down, and strange ones, that are much less massive than heavy quarks as the bottom, charm, and top ones. Heavy quarks can form quarkonia, as the $J/\psi$  meson ground state, and the $\Upupsilon$ mesons as well.

Isovector mesons are characterized by the isobaric spin $I^G=0^-$ with negative $G$-parity, and unit total angular momentum $J^{PC}=1^{--}$, with both parity ($P$) and charge  conjugation ($C$) negative quantum numbers. Isovector mesons   encompass $\Upupsilon$  mesons, consisting of a quarkonium state -- the bottomonium -- formed by a bottom quark ($b$) and its antiparticle ($\bar{b}$);  $\phi$ mesons,  formed of a strange quark and a strange antiquark;  $\psi$  mesons, consisting of a charm quark ($c$) and a charm antiquark ($\bar{c}$) and forming charmonia states;  and $\omega$ mesons as bound states of up and down quarks, with their correspondent antiquarks.
 The top quark has a much bigger mass when compared to the other quarks, being five orders of magnitude more massive than the up and down quarks and two  orders of magnitude more massive than the  bottom and  charm heavy quarks.  Although toponium, formed by a  top quark ($t$) and its antiquark ($\bar{t}$) is more difficult to detect since top quarks decay through the electroweak gauge force before bound states set in, there are recently reported signatures of toponium formation by gluon
fusion, in the LHC run 2 data \cite{Fuks:2021xje}. Contrary to bottomonium and charmonium, toponium resonances decay instantly, as $t$ and $\bar{t}$ have a $5\times10^{-25}$ s lifetime. Toponium has also a huge decay
width.

In this work, the DCE of the isovector mesons   families will be computed, as a function of both the radial $S$-wave resonances excitation level and their experimental mass spectra as well. Hence, the mass spectra of radial $S$-wave resonances with higher excitation levels can be obtained. It is accomplished when plotting the associated graphics and, when interpolating them, two types of DCE Regge  trajectories can be then produced. The first ones relate the DCE to the radial $S$-wave resonances, whereas the second ones associate the DCE with the isovector families experimental mass spectra. Working with the polynomial interpolation formul\ae\, of each interpolation curve for the four families  of isovector mesons ($\Upupsilon$, $\psi$, $\omega$, and $\phi$) yields the mass spectra of higher $n$ resonances in these families. They comprise candidates in the next generations proposed to be detected in experiments, as well as candidates that match up to states in PDG \cite{pdg}.  

This work is organized as follows: Sec. \ref{sec1}  is dedicated  to 
 introducing the basics on AdS/QCD, with a deviation from the soft wall dilaton background that is induced by the massive quarks that constitute isovector mesons. This emulates the standard soft wall when the deviation parameter equals zero and encompasses the light-flavor meson states with massless constituent quarks, in this limit. 
The mass spectra of isovector $\Upupsilon$, $\psi$, $\phi$, and $\omega$ meson resonances are then derived.  
 Sec. \ref{sec2} is hence devoted to the essentials of the DCE setup. The DCE of the isovector $\Upupsilon$, $\psi$, $\phi$, and $\omega$ meson families is calculated as a function of both the radial $S$-wave resonances excitation level and their experimental mass spectra. 
 %The analysis and the interpolation of  the resulting data consist, respectively, of the first and second  types of DCE Regge  trajectories. 
 Extrapolating both DCE Regge  trajectories yields the mass spectra of heavier isovector resonances in each one of the $\Upupsilon$, $\psi$, $\phi$, and $\omega$ meson families, whose first undisclosed states are analyzed and matched up to existing candidates in PDG. In particular, a precise mass spectrum  of bottomonia and charmonia is derived. Sec. \ref{iv} encloses discussion, analysis, and relevant perspectives.

\section{AdS/QCD fundamentals}
\label{sec1}

Isovector mesons are represented, in the soft wall model \cite{Karch:2006pv}, by a vector field  $V_M = (V_\mu,V_z)\,$ where greek indexes run from $0$ to $3$, living in AdS space. 
These fields can be thought of as being dual objects that correspond to the current density 
${\rm J}^\mu(x) = \bar{q}(x)\upgamma^\mu q(x)$ in the CFT on the boundary, where $q(x)$ denote quark fermionic fields and $\upgamma^\mu$ stand for the gamma Dirac matrices, whereas here $\bar{q}(x)\equiv{q}^\dagger(x)\upgamma^0$. 
These  vector field are governed by the action 
\begin{equation}
I \,=- \frac{1}{g_5^2}  \int dz\,{\rm d}^4x \, \sqrt{-g} \,\, \mathcal{L},
\label{vectorfieldaction}
\end{equation}
where \beq\label{lalag}
\mathcal{L}=\frac14\, e^{-\upphi (z)}  F^*_{MN} F^{MN}\eeq
 denotes the Lagrangian and $F_{MN} = \upPartial_MV_N - \upPartial_N V_M$.  
The AdS bulk is equipped with the metric   
\begin{equation}\label{space1}
 ds^2 =g_{MN}dx^Mdx^N= e^{2A(z)}(\upeta_{\mu\nu}dx^\mu dx^\nu+dz^2)\,,
\end{equation}
 with warp factor ${A}(z) = -\log(z/R)$, where $\upeta_{\mu\nu}$ denotes the boundary metric.

 Ref. \cite{MartinContreras:2020cyg} proposed  a deviation  for the soft wall standard quadratic dilaton, 
\beq
\upphi(z) = (\upkappa z)^{2-\upalpha},\label{devia}
\eeq where the energy parameter  $\upkappa$ is   related to the mass scale and $\upalpha$ is the shift parameter driven by constituent massive quarks. Leading $\upalpha$ to  zero, light-flavor mesons with massless constituent quarks are recovered, emulating  the  standard AdS/QCD soft wall.

Using (\ref{vectorfieldaction}) and imposing the gauge  where  $V_z=0$ yields an EOM ruling the bulk field
\begin{equation}
\left[\upPartial_z^2-B'(z)\upPartial_z-q^2\right]V_\mu(z,q)=0,    
\end{equation}
\noindent where $B(z)=\upphi(z)-A(z)$ and the prime denotes the derivative with respect to $z$. The bulk isovector field can be split off as 
\beq\label{qiqi}
V_\mu(z,q)=V_\mu(q)\,\xi(z,q),\eeq where 
   $\xi(z,q)$ denotes the  bulk-to-boundary propagator.    Hence, the EOM ruling the bulk isovector field is led to a Schr\"odinger-like equation. The Bogoliubov mapping $\xi(z,q)=e^{\upphi(z)/2}\upxi(z,q)$ yields  \cite{MartinContreras:2020cyg} 
\begin{equation}\label{schr}
\left[-\upPartial_z^2+U(z)\right]\,\upxi(z,q)=-q^2\upxi(z,q),     
\end{equation}
with the potential 
\begin{eqnarray}\label{potencial11}
U(z)&=&-\frac{1}{2} B''(z)+\frac{1}{4}B^{\prime2}(z).
%&=&\frac{3}{4\,z^2}+\frac{\Phi'(z)}{2\,z}+\frac{\Phi'(z)^2}{4}-\frac{\Phi''(z)}{2},
\end{eqnarray}
\noindent   For the term $V_\mu(q)$, defined in the splitting  \eqref{qiqi} to correspond to the source of the correlators of gauge theory current densities, one must impose the Dirichlet boundary condition $\lim_{z\to0}\xi(z,q) = 1$. 
 The CFT boundary value of the vector field $V_{\mu}(q) \equiv
\lim_{z\to 0} V_\mu (z,q)$ emulates a source that generates correlators of the boundary current operator  $  {\rm J}^\mu (x)$ 
\begin{eqnarray}
\!\! \langle \,0\, \vert \, {\rm J}_\mu (x) {\rm J}_\nu (x') \,  \vert \,0\, \rangle= \frac{\updelta}{\updelta V_{\mu}(x)} \frac{\updelta}{\updelta V_{\nu}(x')}
e^{- I_{\scalebox{.59}{\textsc{on-shell}}}},
 \label{2pointfunction}
\end{eqnarray}
\noindent where $I_{\scalebox{.64}{\textsc{on-shell}}}$  denotes the on-shell action, obtained by the boundary term 
 \begin{equation}\label{lalal}
\!\!\!\!\!\!\!\!I_{\scalebox{.64}{\textsc{on-shell}}}\!=\! - \frac{1}{2 \mathring{g}_5^2}  \lim_{z \to 0}\left[ \int {\rm d}^4x \,\, \frac{1}{z}{e^{-(\upkappa z)^{2-\upalpha}} V_\nu \upPartial_z V^\nu }\right],
\end{equation} for $\mathring{g}_5^2 =  {g_5^2}/{R}$. Eq. \eqref{lalal} generalizes, for the deviated quadratic  dilaton \eqref{devia}, the results that take into account the quadratic dilaton  \cite{Braga:2015jca}. The correlator reads 
 \begin{eqnarray}
  \!\Pi (q^2)  \!=\!   -\frac{1}{\mathring{g}_5^ 2 q^ 2 } \lim_{z \to 0}\left[ \frac{1}{z} e^{ -(\upkappa z)^{2-\upalpha}}    \xi(z,q) \upPartial_z \xi(z,q) 
\right] \,. \label{hol2point}
\end{eqnarray}

The AdS/QCD correspondence asserts that mesonic states can be  established through the conformal dimension $\Updelta$ playing the role of a scaling dimension. The Regge angle for a given mesonic trajectory is defined by the dilaton energy scale $\upkappa$. Bulk fields are dual objects to an  $\mathbb{O}$  operator that is driven by the CFT on the AdS  boundary, whose correlator is given by  $
\left\langle\mathbb{O}(x^\mu)\,\mathbb{O}(\mathbf{0})\right\rangle \propto \|\,x^\mu\,\|^{-2\Updelta}$ \cite{Witten:1998qj}. 
The  spin $S$, the bulk mass, $\mathbb{M}$, and the conformal dimension,  $\Updelta$, are  fine-tuned by the following expression  \cite{Witten:1998qj}, 
\begin{equation}\label{mesao}
    \mathbb{M}_5^2 R^2=(\Updelta-S)(\Updelta+S-4). \end{equation}
Eq. (\ref{mesao}) holds for the $S$-wave, $\ell=0$, approach of isovector  mesons, with $\Updelta =3$ and $S=1$.

Eq. (\ref{schr}) yields the mass spectrum $m_n^2 = -q^2$ of isovector mesons states, with radial excitation level $n$, as  eigenvalues of the Schr\"odinger-like operator. When replacing the  shift of the standard quadratic dilaton \eqref{devia} in the potential \eqref{potencial11} yields  \cite{MartinContreras:2020cyg}
\beq\label{ppp}
\!\!\!\!\!\!\!\!\!\!\!U(z,\upkappa,\upalpha)\!&\!=\!&\!\frac{3}{4 z^2}+\left(\frac{3\upalpha}2-\frac{\upalpha ^2}{2}   -\upkappa ^2\right)\upkappa^2 (\upkappa  z)^{-\upalpha }\nonumber\\
\!&\!+\!&\!\!\frac{\upkappa^2}{4}\left(\upalpha\!-\!2\right)^2\!(\upkappa z)^{2-2 \upalpha }\!+\!\frac{\upkappa }{z}\left(1\!-\!\frac{\upalpha}{2}\right)\!(\upkappa  z)^{1\!-\!\upalpha }.
\eeq
If the quarks that constitute the isovector mesons are taken in the massless limit, corresponding to $\upalpha=0$, the potential (\ref{ppp})  leads to the vector meson fields  \cite{Karch:2006pv}. The parameter  $\upalpha$ encodes the massive quarks constituents and their flavor within isovector mesons \cite{MartinContreras:2020cyg}. Different isovector meson states, with different masses and radial quantum numbers, are distinguished when one adopts, as usual, eigenvalues $-q^2=m_n^2 = 4\kappa^2n$, for $n=1,2,3,\ldots$, corresponding to the mass spectrum derived in Eq. (\ref{schr}). Different radial excitation levels $n$ yield different mesonic excitations in each one of the isovector meson families.

To obtain the radial $S$-wave resonances of the isovector families, the generalized potential in Eq. \eqref{ppp} must be substituted into Eq. (\ref{schr}) , whose numerical solutions, for the isovector meson families,  are encoded in Tables \ref{scalarmasses1} -- \ref{scalarmasses4}, with the  $\upkappa$ and $\upalpha$ parameters 
that are appropriate to fit the mass spectra of each isovector meson family, respectively \cite{MartinContreras:2020cyg}.
It is worth to emphasize that to fit the mass spectra of the four isovector meson families, appropriate values of the two parameters $\upalpha$ and $\upkappa$ must be chosen. For the $\Upupsilon$ meson family, the values $\upalpha= 0.863$ and $\upkappa = 11.209$ GeV best fit the experimental data, whereas $\upalpha= 0.541$ and $\upkappa = 2.150$ GeV are more suitable for describing the family of $\psi$ isovector mesons. Besides, the $\omega$ meson family is best described by adopting  $\upalpha= 0.040$ and $\upkappa = 0.498$ GeV, and to match the mass spectrum of $\phi$ isovector mesons requires $\upalpha= 0.070$ and $\upkappa = 0.585$ GeV.

\begin{table}[H]
\begin{center}-------------- \;$\Upupsilon$ mesons mass spectrum \;---------------\medbreak
\begin{tabular}{||c|c||c|c||}
\hline\hline
$n$ & State & $M_{\scalebox{.67}{\textsc{Experimental}}}$ (MeV)  & $M_{\scalebox{.67}{\textsc{Theory}}}$ (MeV) \\
       \hline\hline
\hline
1 &\;$\Upupsilon(1S)\;$ & $9460.30\pm0.26$ & 9438.51   \\ \hline
2 &\;$\Upupsilon(2S)\;$ & $10023.26 \pm 0.32$ & 9923.32   \\ \hline
3& \;$\Upupsilon(3S)\;$& $10355 \pm 0.5$       & 10277.2     \\\hline
4& \;$\Upupsilon(4S)\!=\!\Upupsilon(10580)\;$& $10579.4 \pm 1.2$  & 10558.6   \\\hline
5& \;$\Upupsilon(10860)\;$& $10889.9^{+3.2}_{-2.6}$     & 10793.5   \\\hline
6& \;$\Upupsilon(11020)\;$&  $11000 \pm 4$   & 10995.7 \\\hline
%7& \;$\eta(\Box)\;$&      &  2087.3  \\\hline
%8& \;$\eta(2225)\;$& $2221^{+13}_{-10}$      & 2193.6   \\\hline
%9& \;$\eta(2320)\;$& $2320\pm 15$      & 2289.4  \\\hline
\hline\hline
\end{tabular}
\caption{Experimental and theoretical mass spectrum of $S$-wave resonances in the $\Upupsilon$ meson family. For the masses in the fourth column, obtained by solving Eq. \eqref{schr}, $\upalpha= 0.863$ and $\upkappa = 11.209$ GeV. } \label{scalarmasses1}
\end{center}
\end{table}
\noindent In particular regarding Table \ref{scalarmasses1}, some issues on the heavier bottomonia, which were detected around a quarter of a century ago, still consist of a conundrum in QCD. Thorough investigations concerning the $\Upupsilon$ meson family, involving the in-depth analysis of available data  provided by CLEO, CUSB, and BaBar collaborations, have been accomplished \cite{pdg,Besson:1984bd,Chen:2016mjn}. It is worth emphasizing that the $\Upupsilon(10580)$ resonance is the  bottomonium 
state of lowest mass, that is above the so-called open-bottom threshold, that decays into two mesons that are composed by a bottom antiquark and either an up, a down, a strange, or a charm quark, originating respectively the $B^+$, $B^0$, $B^0_s$, and $B^+_c$ mesons \cite{Aubert:2004pwa}. 

\begin{table}[H]
\begin{center}-------------- \;$\psi$ mesons mass spectrum \;---------------\medbreak
\begin{tabular}{||c|c||c|c||}
\hline\hline
$n$ & State & $M_{\scalebox{.67}{\textsc{Experimental}}}$ (MeV)  & $M_{\scalebox{.67}{\textsc{Theory}}}$ (MeV) \\
       \hline\hline
\hline
1 &\;$J/\psi(1S)\;$ & $3096.900\pm0.006$ & 3077.09   \\ \hline
2 &\;$\psi(2S)\;$ & $ 3686.10  \pm 0.06$ & 3689.62  \\ \hline
3& \;$\psi(4040) \;$& $4039 \pm 1$       & 4137.5     \\\hline
4& \;$\psi(4415) \;$& $4421\pm 4$  & 4499.4  \\\hline
5& \;$\psi(4660)\;$& $4633 \pm 7 $     & 4806.3    \\\hline
%6& \;$\Upupsilon(11020)\;$&  $11000 \pm 4$   & 10995.7 \\\hline
%7& \;$\eta(\Box)\;$&      &  2087.3  \\\hline
%8& \;$\eta(2225)\;$& $2221^{+13}_{-10}$      & 2193.6   \\\hline
%9& \;$\eta(2320)\;$& $2320\pm 15$      & 2289.4  \\\hline
\hline\hline
\end{tabular}
\caption{Experimental and theoretical mass spectrum of $S$-wave resonances in the $\psi$ meson family. For the masses in the fourth column, obtained by solving Eq. \eqref{schr}, $\upalpha= 0.541$ and $\upkappa = 2.150$ GeV.} \label{scalarmasses2}
\end{center}
\end{table}
Regarding Tables \ref{scalarmasses2}
 and \ref{scalarmasses3},  Belle, CDF,  and
LHCb collaborations produced relevant data, in particular about the $\omega$ resonances, whose narrow width can induce relevant effects of scattering involving the $J/\psi$, $\omega$, and charmonium \cite{Braaten:2013poa}.  
\begin{table}[H]
\begin{center}-------------- \;$\omega$ mesons mass spectrum \;---------------\medbreak
\begin{tabular}{||c|c||c|c||}
\hline\hline
$n$ & State & $M_{\scalebox{.67}{\textsc{Experimental}}}$ (MeV)  & $M_{\scalebox{.67}{\textsc{Theory}}}$ (MeV) \\
       \hline\hline
\hline
1 &\;$\omega(782)\;$ & $782.65\pm0.12$ & 981.43   \\ \hline
2 &\;$\omega(1420)\;$ & $1410 \pm 60$ & 1374  \\ \hline
3& \;$\omega(1650) \;$& $1670\pm 30$       & 1674     \\\hline
4& \;$\omega(1960) \;$& $1960\pm25$  & 1967  \\\hline
5& \;$\omega(2205) \;$& $2205 \pm 30 $     & 2149   \\\hline
6& \;$\omega(2290)\;$&  $2290 \pm 20$   & 2348 \\\hline
%7& \;$\eta(\Box)\;$&      &  2087.3  \\\hline
%8& \;$\eta(2225)\;$& $2221^{+13}_{-10}$      & 2193.6   \\\hline
%9& \;$\eta(2320)\;$& $2320\pm 15$      & 2289.4  \\\hline
\hline\hline
\end{tabular}
\caption{Experimental and theoretical mass spectrum of $S$-wave resonances in the $\omega$ meson family. For the masses in the fourth column, obtained by  solving Eq. \eqref{schr}, $\upalpha= 0.040$ and $\upkappa = 0.498$ GeV.} \label{scalarmasses3}
\end{center}
\end{table}

\begin{figure}[H]
	\centering
	\includegraphics[width=8.5cm]{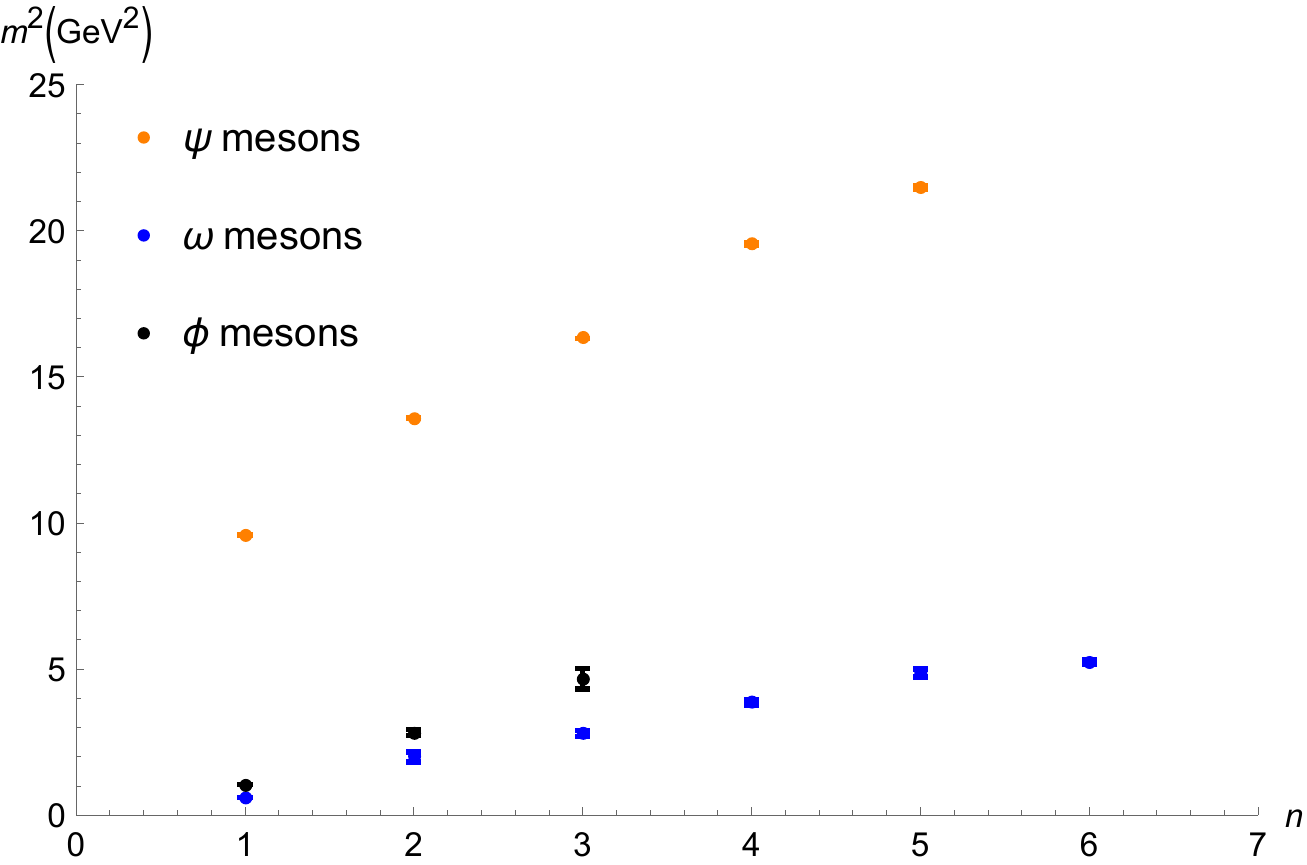}
	\caption{Experimental squared mass spectrum of the isovector meson families, with respect to the $S$-wave resonances excitation levels, $n$, with error bars. The orange points correspond to the $\psi$ meson family, for  $n=1,\ldots,5$, respectively coinciding to the $J/\psi(1S)$, $\psi(2S)$, $\psi(4040)$, $\psi(4415)$, and $\psi(4660)$ $S$-wave resonances;  the blue points regards the $\omega$ meson family, for  $n=1,\ldots,6$, respectively corresponding to the $\omega(782)$, $\omega(1420)$, $\omega(1650)$, $\omega(1960)$, $\omega(2205)$, and $\omega(2290)$  $S$-wave resonances; and the black points depict the $\phi$ meson family, for  $n=1,\ldots,3$, respectively corresponding to the $\phi(1020)$, $\phi(1680)$, and $\phi(2170)$ 
 $S$-wave resonances in PDG \cite{pdg}.}
	\label{tudonmn}
\end{figure}

\begin{figure}[H]
	\centering
	\includegraphics[width=8.3cm]{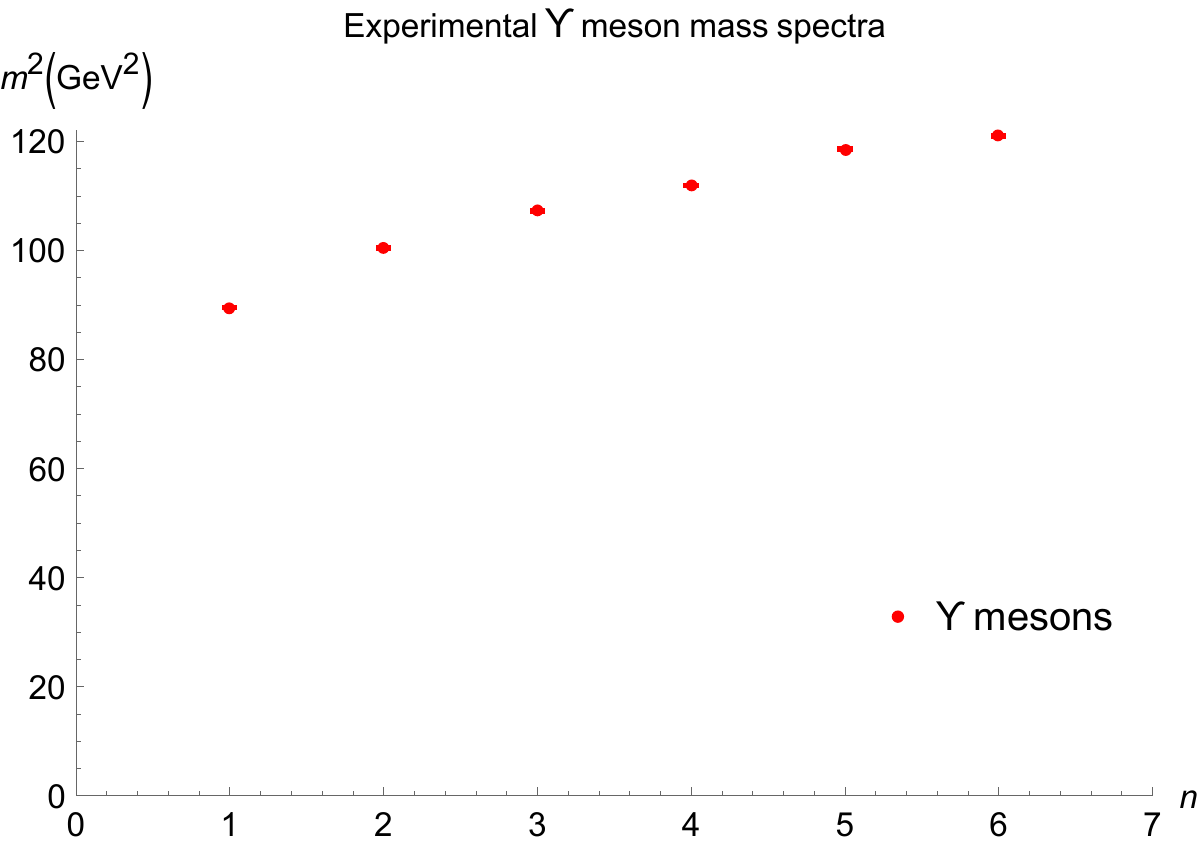}
	\caption{Experimental squared mass spectrum of the $\Upupsilon$ family, for $n=1,\ldots,6$, with error bars, respectively corresponding to the $\Upupsilon(1S)$, $\Upupsilon(2S)$, $\Upupsilon(3S)$, $\Upupsilon(4S)=\Upupsilon(10580)$, $\Upupsilon(10860)$, and $\Upupsilon(11020)$ $S$-wave resonances in PDG \cite{pdg}.}
	\label{upsilonnmn}
\end{figure}

\begin{table}[H]
\begin{center}-------------- \;$\phi$ mesons mass spectrum \;---------------\medbreak
\begin{tabular}{||c|c||c|c||}
\hline\hline
$n$ & State & $M_{\scalebox{.67}{\textsc{Experimental}}}$ (MeV)  & $M_{\scalebox{.67}{\textsc{Theory}}}$ (MeV) \\
       \hline\hline
\hline
1 &\;$\phi(1020)\;$ & $1019.461\pm0.016$ & 1139.43   \\ \hline
2 &\;$\phi(1680)\;$ & $1680 \pm 20$ & 1583  \\ \hline
%3& \;$\phi(1650) \;$& $1670\pm 30$       & 1921     \\\hline
3& \;$\phi(2170) \;$& $2160\pm80$  & 2204  \\\hline
%5& \;$\omega(2205) \;$& $2205 \pm 30 $     & 2149   \\\hline
%6& \;$\omega(2290)\;$&  $2290 \pm 20$   & 2348 \\\hline
%7& \;$\eta(\Box)\;$&      &  2087.3  \\\hline
%8& \;$\eta(2225)\;$& $2221^{+13}_{-10}$      & 2193.6   \\\hline
%9& \;$\eta(2320)\;$& $2320\pm 15$      & 2289.4  \\\hline
\hline\hline
\end{tabular}
\caption{Experimental and theoretical mass spectrum of the $S$-wave resonances in the $\phi$ meson family. For the masses in the fourth column, obtained by solving Eq. \eqref{schr}, $\upalpha= 0.070$ and $\upkappa = 0.585$ GeV. } \label{scalarmasses4}
\end{center}
\end{table}
Data in Tables \ref{scalarmasses1} -- \ref{scalarmasses4} are displayed in Figs. \ref{tudonmn} and \ref{upsilonnmn}.

\section{Mass spectroscopy of isovector mesons from DCE}
\label{sec2}
The DCE  quantifies the correlations out of the oscillations of energy configurations associated with physical systems under scrutiny. For it, the energy density, namely the temporal $\tau_{00}({\bf r})$ component of the stress-energy tensor field, for ${\bf r}=(x_1,\ldots,x_k)\in\mathbb{R}^k$, emulates a measurable Lebesgue-integrable localized scalar field. The correlator 
\beq
\Uppi({\bf r})=\int_{\mathbb{R}^k} {\rm d}^k x\, \tau_{00}(\mathring{\bf r})\tau_{00}(\mathring{\bf r}+{\bf r})\eeq
 sets the DCE to be thought of as the Shannon's information  entropy that regulates the correlations among the system constituents \cite{Braga:2018fyc}.  
 Now denoting by ${\bf q}$ the spatial part of the 4-momentum $q$, the protocol to derive the DCE has a first step of computing the  Fourier transform,  
\beq\label{fou}
\tau_{00}({\bf q}) = \frac{1}{(2\pi)^{k/2}}\int_{\mathbb{R}^k}{\rm d}^k x\,\tau_{00}({\bf r})e^{-i{\bf q}\cdot {\bf r}}.\eeq 
Any given wave mode enclosed by a $k$-volume ${\rm d}^k{q}$, which has  center at ${\bf q}$, is related to a probability, $\mathcal{P}$, corresponding to the spectral density in the given wave 
mode, that reads \cite{Gleiser:2018kbq}
\beq
\mathcal{P}\left({\bf q}\,\vert\, {\rm d}^kq\right)\propto \left|\tau_{00}({\bf q})\right|^{2}{\rm d}^kq.
\eeq     The  probability distribution (\ref{fou}) engenders the  modal fraction     
\cite{Gleiser:2012tu}, 
\begin{eqnarray}
\uptau_{00}({\bf q}) = \frac{\left|\tau_{00}({\bf q})\right|^{2}}{ \int_{\mathbb{R}^k} \,{\rm d}^k{q}\, \left|\tau_{00}({\bf q})\right|^{2}}.\label{modalf}
\end{eqnarray} 
The weight of information that is necessary to encode $\tau_{00}$, with respect to wave modes, is calculated by the DCE, 
\begin{eqnarray}
{\rm DCE}_{\tau_{00}}= - \int_{\mathbb{R}^k}\,{\rm d}^kq\,{\uptau_{00}^{\scalebox{.73}{$\,\triangleright$}}}({\bf q})\log  {\uptau_{00}^{\scalebox{.73}{$\,\triangleright$}}}({\bf q})\,,
\label{confige}
\end{eqnarray}
where {$\uptau_{{00}}^{\scalebox{.73}{$\,\triangleright$}}({\bf q})=\uptau_{00}({\bf q})/\uptau_{{00}}^{\scalebox{.68}{sup}}({\bf q})$}, and ${\uptau_{{00}}^{\scalebox{.68}{sup}}({\bf q})}$ denotes the supremum value of {$\uptau_{00}$} attained in $\mathbb{R}^k$. Besides, in the state ${\uptau_{{00}}^{\scalebox{.68}{sup}}({\bf q})}$, the spectral density reaches its highest value. The DCE is measured in terms of nat/unit volume. The acronym \emph{nat} designates the information natural unit, where nat$^{-1}= \log 2$/bits. It is equivalent to the information of a uniform probability distribution in the interval of the real line between the origin and the Euler's number.  The DCE encrypts the information of scale, since the spectral density is represented by the Fourier transform of the correlator \cite{Gleiser:2018kbq}.

To compute the DCE of isovector families, $k=1$ is chosen in Eqs. (\ref{fou}) -- (\ref{confige}),  since the CFT boundary has codimension one. Replacing the Lagrangian (\ref{lalag}) into the stress-energy tensor field expression,  
 \begin{equation}
{ \!\!\!\!\!\!\!\!\tau_{00}\!=\!  \frac{2}{\sqrt{ -g }}\!\! \left[\frac{\upPartial (\sqrt{-g}{\mathcal{L}})}{\upPartial{g^{00}}} \!-\!\frac{\upPartial}{\upPartial{ x^\beta }}  \frac{\upPartial (\sqrt{-g} {\mathcal{L}})}{\upPartial\left(\frac{{\scalebox{.79}{$\,\upPartial$}} g^{00}}{{\scalebox{.79}{$\,\upPartial$}}x^\beta}\right)}
%  \!+\!\mathcal{T}^{mn}\!
  \right]},
  \label{em1}
 \end{equation} 
 %\noindent  
and subsequently computing its Fourier transform, the modal fraction, and the DCE, respectively using Eqs. (\ref{fou}) -- (\ref{confige}), we can derive the mass spectra of the isovector meson families.
The energy density can be then written as
\beq\label{t001}
\!\!\!\!\!\!\tau_{00}\!=\!f(\upkappa,\upalpha)g_{00}\left(g^{MP}g^{NQ}F_{MN}F_{PQ}\right)\!-\!g^{MN}F_{0M}F_{0N},
\eeq
for $P,Q=0,1,2,3,5$, denoting $f(\upkappa,\upalpha)={e^{-(\upkappa z)^{2-\upalpha}}}/{4g_5^2}$. 
Now when a plane wave solution is considered in the isovector meson rest frame $V_\mu = \epsilon_\mu \xi(q, z)e^{-imc^2t}$, without loss of generality, the polarization $\epsilon_\mu = (0, 1, 0, 0)^\intercal$ yields 
\beq\label{t002}
\!\!\!\!\!\tau_{00}=\frac{z^2e^{-(\upkappa z)^{2-\upalpha}}}{2g_5^2R^2}\left[-\left(\upPartial_z\xi\right)^{2}+m_n^2\xi^2\right],
\eeq again stressing the fact that $m_n^2=4\kappa^2n$, for $n=1,2,3,\ldots$, determines the mass spectra for different isovector meson states.

 This technique, taking into account the information content of AdS/QCD,  based on the interpolation of the experimental mass spectra of the four families of isovector mesons in PDG \cite{pdg}, is more precise when compared to just solving Eq. (\ref{schr}) to derive the mass spectra. The DCE, using the protocol (\ref{fou} -- \ref{confige}), for each isovector  meson family, is numerically calculated and exhibited in  Tables \ref{scalarmasses5} -- \ref{scalarmasses8}. 
\begin{table}[h!]
\begin{center}
\begin{tabular}{||c|c|c||}
\hline\hline
$n$ & State & DCE (nat) \\
       \hline\hline
\hline
1 &\;$\Upupsilon(1S)\;$ & 24.78  \\ \hline
2 &\;$\Upupsilon(2S)\;$ &  26.71  \\ \hline
3& \;$\Upupsilon(3S)\;$&  28.30  \\\hline
4& \;$\Upupsilon(4S)=\Upupsilon(10580)\;$& 29.88 \\\hline
5& \;$\Upupsilon(10860)\;$& 31.92  \\\hline
6& \;$\Upupsilon(11020)\;$& 33.96 \\\hline
\hline\hline
\end{tabular}
\caption{DCE of the $S$-wave resonances in the $\Upupsilon$ meson family.} \label{scalarmasses5}
\end{center}
\end{table}

\begin{table}[H]
\begin{center}
\begin{tabular}{||c|c|c||}
\hline\hline
$n$ & State & DCE (nat) \\
       \hline\hline
\hline
1 &\;$J/\psi(1S)\;$ &  19.45 \\ \hline
2 &\;$\psi(2S)\;$ & 20.81  \\ \hline
3& \;$\psi(4040) \;$&   22.07  \\\hline
4& \;$\psi(4415) \;$&  23.97 \\\hline
5& \;$\psi(4660)\;$&  25.64   \\\hline
\hline\hline
\end{tabular}
\caption{DCE of the $S$-wave resonances in the $\psi$ meson family.} \label{scalarmasses6}
\end{center}
\end{table}

\begin{table}[H]
\begin{center}
\begin{tabular}{||c|c|c||}
\hline\hline
$n$ & State & DCE (nat) \\
       \hline\hline
\hline
1 &\;$\omega(782)\;$ & 11.37  \\ \hline
2 &\;$\omega(1420)\;$ & 12.46  \\ \hline
3& \;$\omega(1650) \;$& 13.51    \\\hline
4& \;$\omega(1960) \;$& 15.14 \\\hline
5& \;$\omega(2205) \;$& 17.36   \\\hline
6& \;$\omega(2290)\;$&  18.69 \\\hline
%7& \;$\eta(\Box)\;$&      &  2087.3  \\\hline
%8& \;$\eta(2225)\;$& $2221^{+13}_{-10}$      & 2193.6   \\\hline
%9& \;$\eta(2320)\;$& $2320\pm 15$      & 2289.4  \\\hline
\hline\hline
\end{tabular}
\caption{DCE of the $S$-wave resonances in the $\omega$ meson family.} \label{scalarmasses7}
\end{center}
\end{table}

\begin{table}[H]
\begin{center}
\begin{tabular}{||c|c|c||}
\hline\hline
$n$ & State &DCE (nat) \\
       \hline\hline
\hline
1 &\;$\phi(1020)\;$ &  13.48  \\ \hline
2 &\;$\phi(1680)\;$ & 15.39  \\ \hline
3& \;$\phi(2170) \;$& 17.52  \\\hline
%5& \;$\omega(2205) \;$& $2205 \pm 30 $     & 2149   \\\hline
%6& \;$\omega(2290)\;$&  $2290 \pm 20$   & 2348 \\\hline
%7& \;$\eta(\Box)\;$&      &  2087.3  \\\hline
%8& \;$\eta(2225)\;$& $2221^{+13}_{-10}$      & 2193.6   \\\hline
%9& \;$\eta(2320)\;$& $2320\pm 15$      & 2289.4  \\\hline
\hline\hline
\end{tabular}
\caption{DCE of the $S$-wave resonances in the $\phi$ meson family. } \label{scalarmasses8}
\end{center}
\end{table}

\begin{figure}[H]
	\centering
	\includegraphics[width=8.9cm]{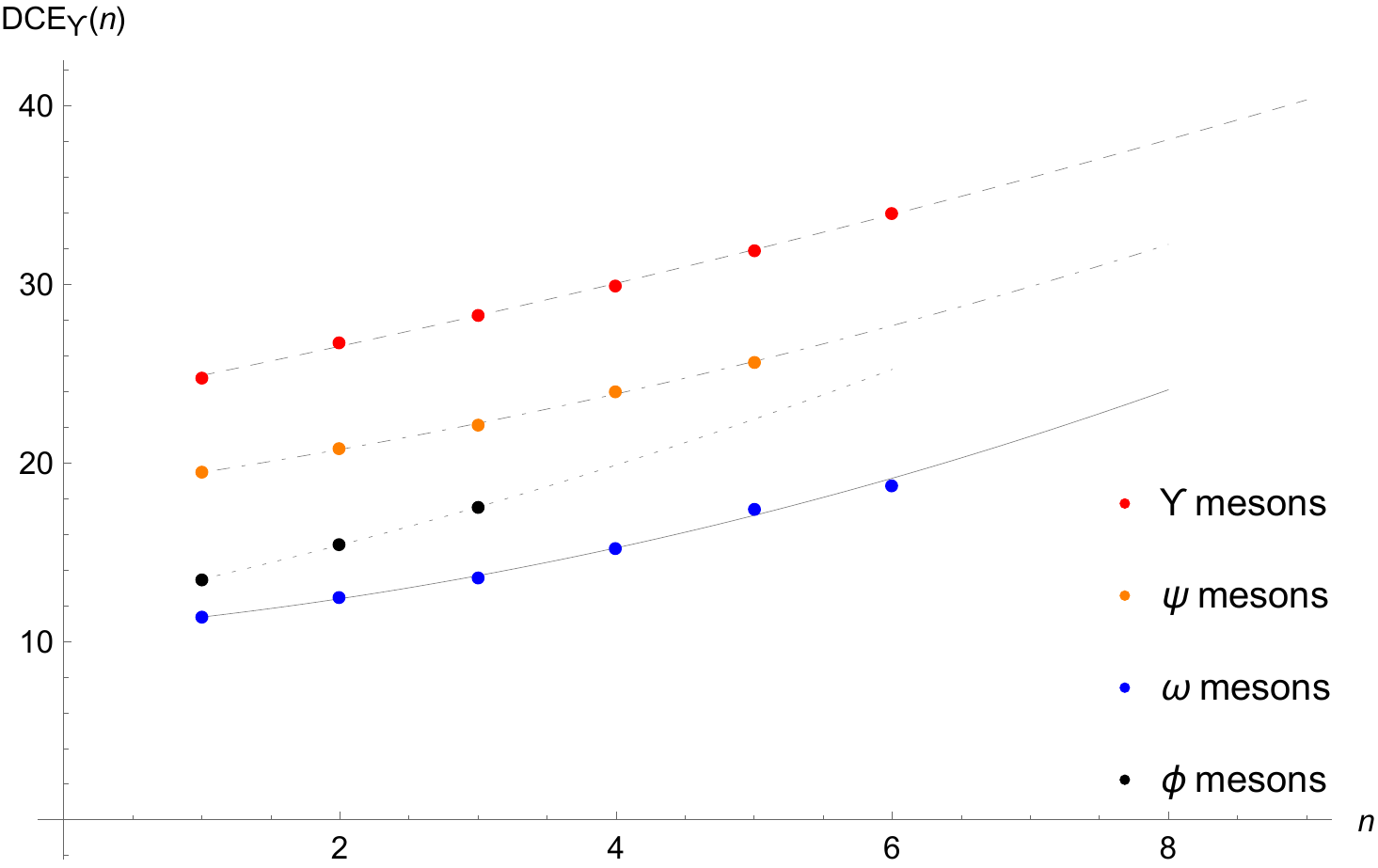}
	\caption{DCE of the isovector meson families as a function of the $n$ radial excitation level.  
The first form  of DCE Regge  trajectory is plotted as the   dashed line, for the $\Upupsilon$ mesons; as the dot-dashed line, for the $\psi$ mesons; as the dotted line, for the $\phi$ mesons; and as the gray line for the $\omega$ mesons.}
	\label{cen1}
\end{figure}

The  first form of DCE Regge  trajectories regards the DCE with respect to the $n$ radial excitation level of isovector meson resonances. Fig. \ref{cen1} shows the obtained outcomes, whose cubic polynomial  interpolation of data in Tables \ref{scalarmasses5} -- \ref{scalarmasses8} respectively generates the first type of DCE Regge  trajectories, 
\begin{subequations}
\begin{eqnarray}\label{itp1}
\!\!\!\!\!\!\!\!\!\!\!\!\clt{{\rm DCE}_\Upupsilon(n)}\!\!&\!=\!&\clt{4.196\times 10^{-2}n^2+1.509n+ 23.339},
\\
\!\!\!\!\!\!\!\!\! \!\!\!\!\clt{{\rm DCE}_\psi(n)} \label{itp2}\!\!&\!=&\clt{0.090 n^2+1.014n+18.356},\\
\!\!\!\!\!\!\!\!\! \!\!\!\!\clt{{\rm DCE}_\omega(n)} \label{itp3}\!\!&\!=\!&\clt{0.132 n^2+0.632n+10.591},\\
\!\!\!\! \!\!\!\!\! \!\!\!\!\clt{{\rm DCE}_\phi(n)}\label{itp4} \!\!&\!=\!&\! \clt{0.110n^2+1.580n+11.790},   \end{eqnarray}
   \end{subequations}
 \clt{within $1\%$ ($\Upupsilon$ and $\psi$), $1.2\%$ ($\omega$), and $0.2\%$ ($\phi$) root-mean-square deviation (RMSD)}.

\begin{figure}[H]
	\centering
	\includegraphics[width=8.5cm]{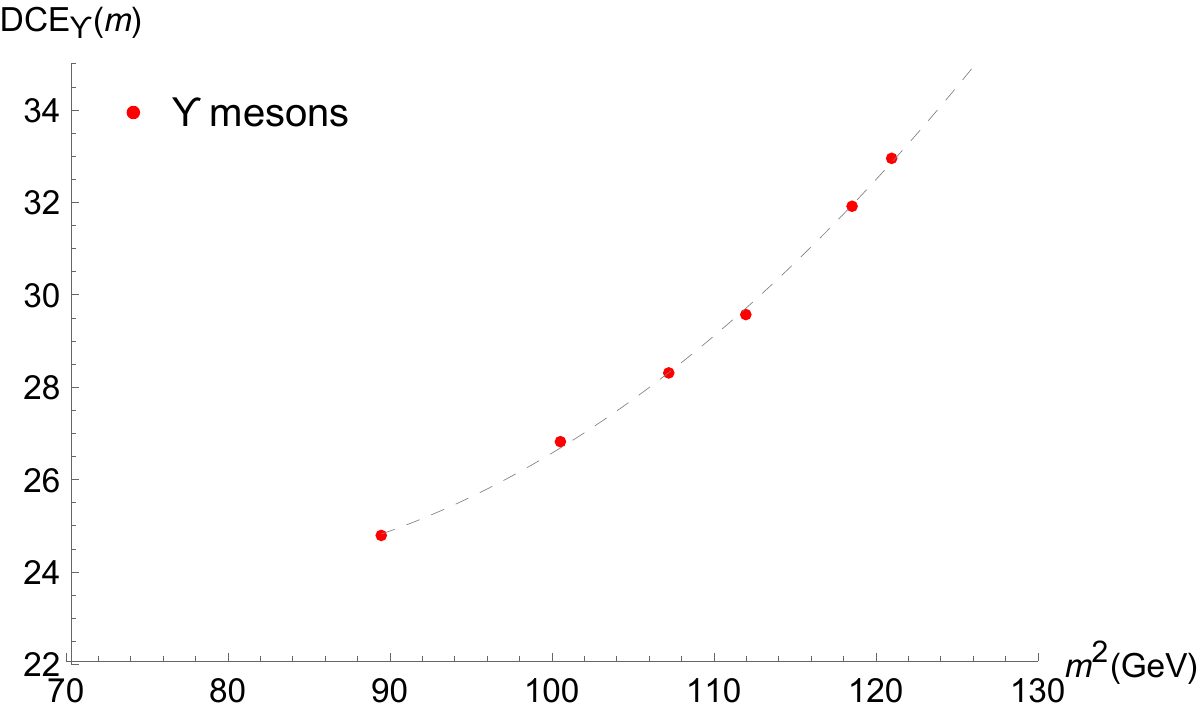}
	\caption{DCE of the $\Upupsilon$ meson family, for  $n=1,\ldots,6$ (respectively corresponding to the $\Upupsilon(1S)$, $\Upupsilon(2S)$, $\Upupsilon(3S)$, $\Upupsilon(4S)=\Upupsilon(10580)$, $\Upupsilon(10860)$ and $\Upupsilon(11020)$ $S$-wave resonances in PDG \cite{pdg}) with respect to their squared mass. 
The DCE Regge  trajectory is represented by the interpolating dashed line.}
	\label{cem11}
\end{figure}

\begin{figure}[H]
	\centering
	\includegraphics[width=8.5cm]{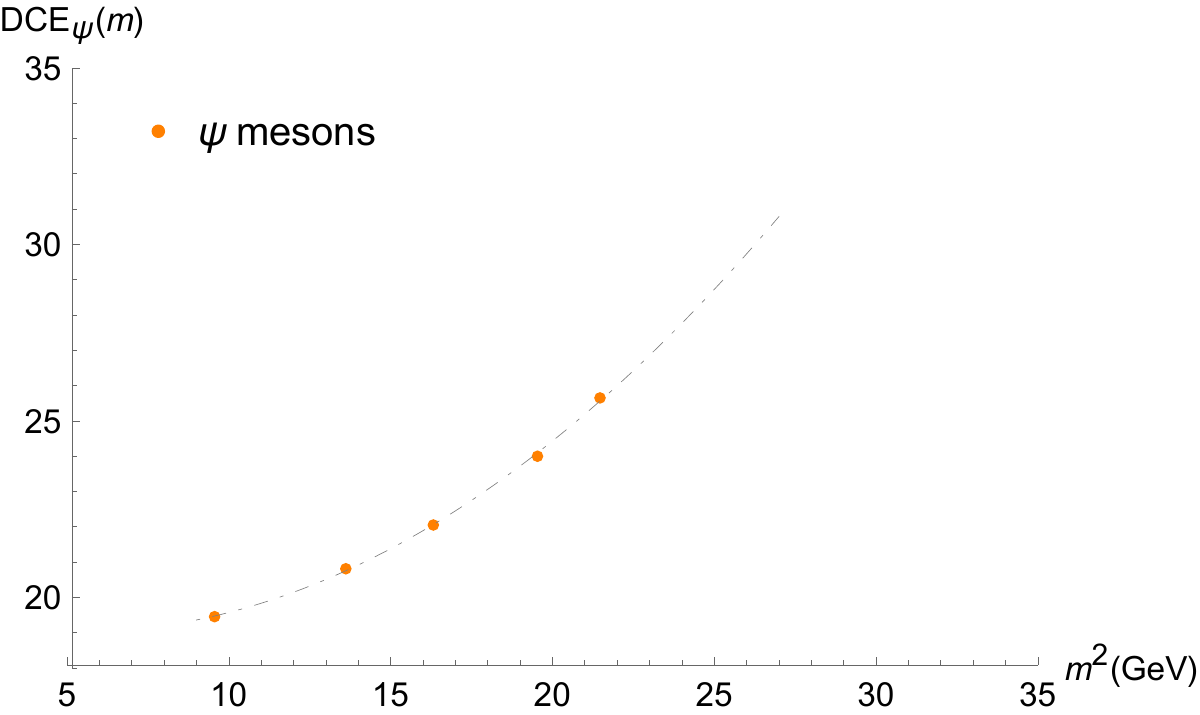}
	\caption{DCE of the $\psi$ meson family, for  $n=1,\ldots,5$ (respectively corresponding to the $J/\psi(1S)$, $\psi(2S)$, $\psi(4040)$, $\psi(4415)$, and $\psi(4660)$ $S$-wave resonances in PDG \cite{pdg}) with respect to their squared mass. 
The DCE Regge  trajectory is represented by the interpolating dot-dashed line.}
	\label{cem12}
\end{figure}

\begin{figure}[H]
	\centering
	\includegraphics[width=8.5cm]{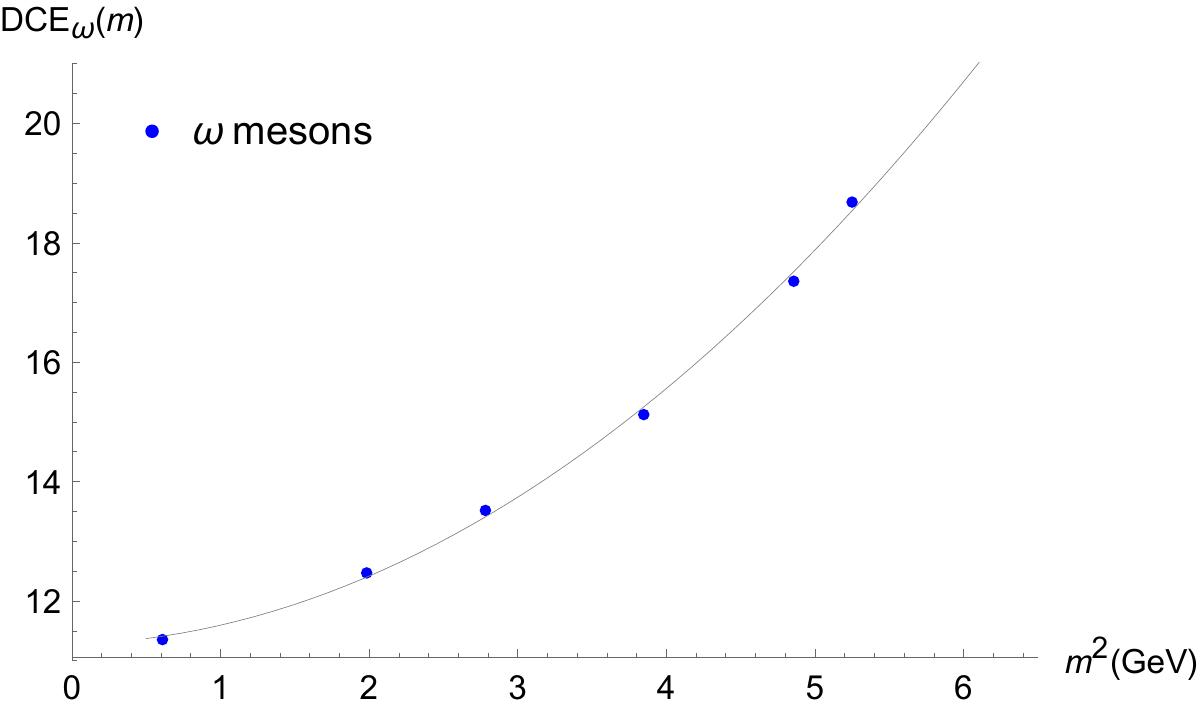}
	\caption{DCE of the $\omega$ meson family, for  $n=1,\ldots,6$ (respectively corresponding to the $\omega(782)$, $\omega(1420)$, $\omega(1650)$, $\omega(1960)$, $\omega(2205)$, and $\omega(2290)$  $S$-wave resonances in PDG \cite{pdg}) with respect to their squared mass. 
The DCE Regge  trajectory is represented by the interpolating gray line.}
	\label{cem13}
\end{figure}

\begin{figure}[H]
	\centering
	\includegraphics[width=8.5cm]{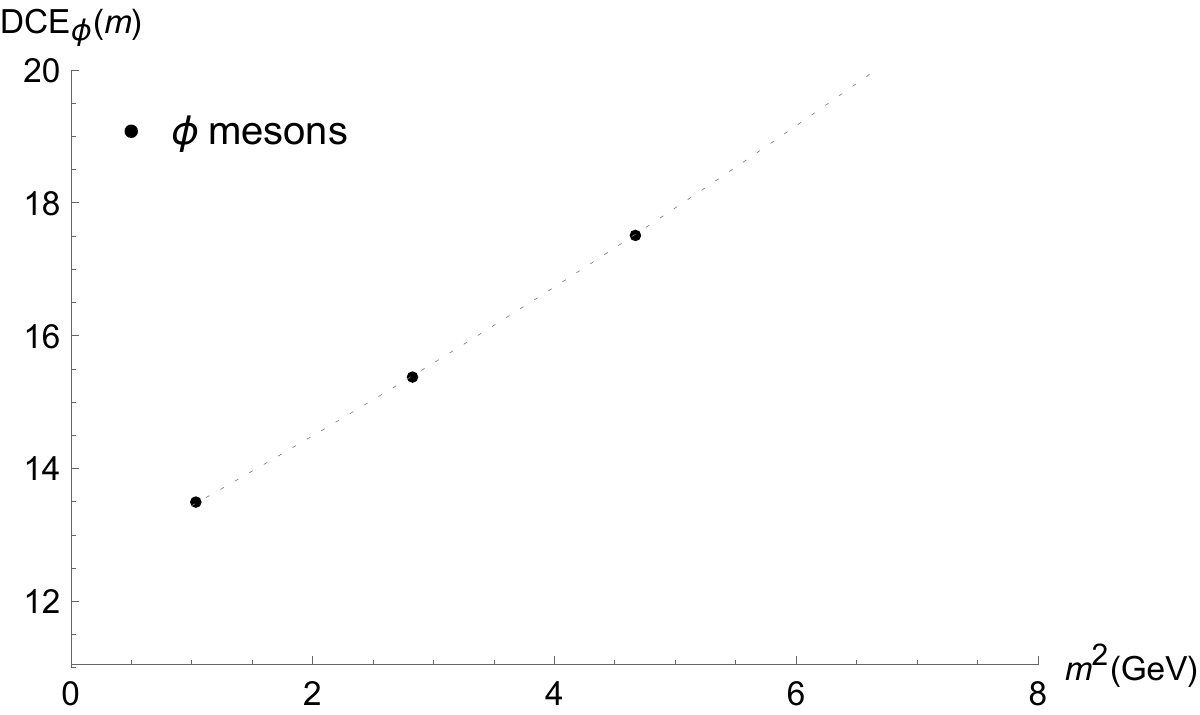}
	\caption{DCE of the $\phi$ meson family, for  $n=1,\ldots,3$ (respectively corresponding to the $\phi(1020)$, $\phi(1680)$, and $\phi(2170)$ 
 $S$-wave resonances in PDG \cite{pdg}) with respect to their squared mass. 
The DCE Regge  trajectory is represented by the interpolating dotted line.}
	\label{cem14}
\end{figure}

Linear Regge trajectories illustrate the proportionality between the radial excitation level of light-flavor mesons and the square of their mass spectrum. However, it does not hold necessarily, when heavy quarks constituents are taken into account. The DCE of isovector meson families can be  also realized  as a function of mass spectra of the isovector meson resonances. In this way, the second type of DCE Regge  trajectories regard the experimentally detected  mass spectra of the $\Upupsilon$, $\psi$, $\omega$, and $\phi$ meson $S$-wave resonances \cite{pdg}. 
In fact, having the DCE of all states in the four isovector meson families in Tables \ref{scalarmasses5} -- \ref{scalarmasses8}, we can also plot the DCE as a function of the mass of each resonance, available in Tables \ref{scalarmasses1} -- \ref{scalarmasses4}. The results are shown in Figs. \ref{cem11} -- \ref{cem14}, respectively, whose interpolation method generates the second type of DCE Regge trajectories in Eqs. (\ref{itq11}) -- (\ref{itq14}). These equations are the second type of DCE Regge trajectories, respectively, for the $\Upupsilon$, $\psi$, $\omega$, and $\phi$ isovector meson families.

\noindent The second type of DCE Regge  trajectories, relating the DCE of the isovector mesons to their squared mass spectra, $m^2$ (GeV${}^2$), respectively for the $\Upupsilon$, $\psi$, $\omega$, and $\phi$ meson families, read
\begin{subequations}
\begin{eqnarray}
\label{itq11}
\!\!\!\!\!\!\!\!\!\!\!\!\!\!\!\!\!\!\clt{{\rm DCE}_\Upupsilon(m)} \!&\!=\!& \clt{4.242\times 10^{-3} m^4\!-\!0.637 m^2\!+\!47.892},\\
 \label{itq12}
\!\!\!\!\!\!\!\!\!\!\!\!\!\!\!\!\!\!\clt{ {\rm DCE}_\psi(m)} \!&\!=\!&\clt{2.487\times10^{-2}m^4 0.260m^2+19.688},\\
 \label{itq13}
\!\!\!\!\!\!\!\!\!\!\!\!\!\!\!\!\!\!\clt{{\rm DCE}_\omega(m)} \!&\!=\!&\clt{0.249 m^4+0.073m^2 +11.273},\\
 \label{itq14}
\!\!\!\!\!\!\!\!\!\!\!\!\!\!\!\!\!\!\clt{{\rm DCE}_\phi(m)} \!&\!=\!&  \clt{2.328 \times10^{-2}m^4 \!+\!0.981m^2\!+\!12.435},
   \end{eqnarray}\end{subequations} \clt{within $1\%$ ($\Upupsilon$ and $\psi$), $1.3\%$ ($\omega$), and $0.1\%$ ($\phi$) root-mean-square deviation (RMSD)}.    
   
Eqs. (\ref{itp1}) -- (\ref{itp4}) and (\ref{itq11}) -- (\ref{itq14}), pairwise respectively, carry the core of properties of the isovector  meson families. Besides, from Eqs. (\ref{itp1}) -- (\ref{itp4}) one can promptly infer the DCE of elements in each family for higher values of the $n$ excitation level of $S$-wave resonances. Subsequently, with the DCE in hands for each $n$ excitation level, one can substitute it on the left-hand side of Eqs. (\ref{itq11}) -- (\ref{itq14}) and solve them,  deriving the mass of each isovector meson resonance corresponding to a higher $n$ excitation level. In this way, the mass spectrum of new elements in each isovector meson family can be obtained. Besides, this method is based solely on the experimental mass spectra of the isovector meson families, which is a more realistic technique -- in the sense that it uses the experimental mass spectra of isovector families --  when compared to pure AdS/QCD predictions. Here we use the stress-energy tensor of AdS/QCD to compute the DCE according to (\ref{fou}) -- (\ref{confige}), however Eqs.  (\ref{itq11}) -- (\ref{itq14}) are obtained by interpolation of the experimental mass spectra of the isovector meson families, also shown in Figs. \ref{cem11} -- \ref{cem14}.

 Let us begin the analysis  with the $\Upupsilon$ meson family.
 We want to determine the mass of the  $\Upupsilon(7S)$ element in this meson family, corresponding to the $n=7$  resonance. 
When  putting $n=7$ back into Eq. (\ref{itp1}), the DCE amounts to \clt{35.961 nat}. Afterwards, using this value of DCE into the left hand side of Eq. (\ref{itq11}), one can obtain the  algebraic solution for the $m$ variable, yielding the mass of the  $\Upupsilon(7S)$ resonance to be  \clt{$m_{\Upupsilon(7S)}= 11.328$ GeV}. Now, accomplishing the same technique to the $n=8$ excitation level in Eq. (\ref{itp1}), the DCE underlying the $\Upupsilon(8S)$ resonance reads \clt{38.099 nat}. Subsequently  working out Eq. (\ref{itq11}) for this value of the DCE implies the $\Upupsilon(8S)$ resonance mass to assume the value  \clt{$m_{\Upupsilon(8S)}= 11.527$ GeV}.  
Analogously,  the $n=9$ excitation level can be now implemented into Eq. (\ref{itp1}), implying the DCE of the $\Upupsilon(9S)$ resonance to be equal to \clt{40.322 nat}. When Eq. (\ref{itq11}) is then resolved, using this value of DCE, the $\Upupsilon(9S)$ resonance mass  \clt{$m_{\Upupsilon(9S)}= 11.715$ GeV} is acquired. These results are compiled in Table 
\ref{scalarmasses10}. 
	\begin{table}[H]
\begin{center}\begin{tabular}{||c|c||c|c||}
\hline\hline
$n$ & State & $M_{\scalebox{.67}{\textsc{Experimental}}}$ (MeV)  & $M_{\scalebox{.67}{\textsc{Theory}}}$ (MeV) \\
       \hline\hline
\hline
1 &\;$\Upupsilon(1S)\;$ & $9460.30\pm0.26$ & 9438.51   \\ \hline
2 &\;$\Upupsilon(2S)\;$ & $10023.26 \pm 0.32$ & 9923.32   \\ \hline
3& \;$\Upupsilon(3S)\;$& $10355 \pm 0.5$       & 10277.2     \\\hline
4& \;$\Upupsilon(4S)\!=\!\Upupsilon(10580)\;$& $10579.4 \pm 1.2$  & 10558.6   \\\hline
5& \;$\Upupsilon(10860)\;$& $10889.9^{+3.2}_{-2.6}$     & 10793.5   \\\hline
6& \;$\Upupsilon(11020)\;$&  $11000 \pm 4$   & 10995.7 \\\hline
7*& \;$\Upupsilon(7S)\;$&    -----  &  \clt{11328.1${}^\star$} \\\hline
8*& \;$\Upupsilon(8S)\;$& -----      & \clt{11527.5${}^\star$}   \\\hline
9*& \;$\Upupsilon(9S)\;$& -----     & \clt{11715.2${}^\star$} \\\hline
\hline\hline
\end{tabular}
\caption{Table \ref{scalarmasses1} completed with the higher $n$  resonances (with an asterisk) of the $\Upupsilon$ meson family. The extrapolated masses for $n=7, 8, 9$ in the fourth column indicated with a `` ${}^\star$ '' specify that they are result of the concomitant use of DCE Regge  trajectories (\ref{itp1}, \ref{itq11}), interpolating the experimental masses of $\Upupsilon$ mesons from $n=1,\ldots,6$. } \label{scalarmasses10}
\end{center}
\end{table}

Now, let us use the DCE Regge trajectories (\ref{itp2}, \ref{itq12}) for scrutinizing the $\psi$ meson family. Firstly, we want to determine the mass of the $\psi(6S)$ meson resonance, that corresponds to the $n=6$ radial excitation level. Putting $n=6$ back into Eq. (\ref{itp2}) implies that the DCE is equal to \clt{27.680 nat}. Therefore, by replacing it in the left-hand side of Eq. (\ref{itq12}), one can solve the algebraic equation for the variable $m$. It yields the mass of the $\psi(6S)$ meson resonance to be equal to \clt{$m_{\psi(6S)}= 4889.5$ MeV}. Carrying out the same protocol for the $n=7$ excitation level, using the DCE Regge  trajectory (\ref{itp2}) implies that the DCE of the $\psi(7S)$ resonance is equivalent to  \clt{29.864 nat}. Hence when solving Eq. (\ref{itq12}), once this last value of DCE is substituted into its left-hand side, the mass of the $\psi(7S)$ meson resonance reads \clt{$m_{\psi(7S)}= 5111.4$ MeV}. 
In the same way,  now utilizing the $n=8$ excitation level into Eq. (\ref{itp2}) yields the DCE of the $\psi(8S)$ resonance to be equal to \clt{32.228 nat}. In this case, the DCE Regge  trajectory  (\ref{itq12}) can be solved for this value, implying that the mass of the $\psi(8S)$ meson is given by \clt{$m_{\psi(8S)}= 5318.7$ MeV}. These results are illustrated in Table 
\ref{scalarmasses20}.

\begin{table}[H]
\begin{center}
\begin{tabular}{||c|c||c|c||}
\hline\hline
$n$ & State & $M_{\scalebox{.67}{\textsc{Experimental}}}$ (MeV)  & $M_{\scalebox{.67}{\textsc{Theory}}}$ (MeV) \\
       \hline\hline
\hline
1 &\;$J/\psi(1S)\;$ & $3096.900\pm0.006$ & 3077.09   \\ \hline
2 &\;$\psi(2S)\;$ & $ 3686.10  \pm 0.06$ & 3689.62  \\ \hline
3& \;$\psi(4040) \;$& $4039 \pm 1$       & 4137.5     \\\hline
4& \;$\psi(4415) \;$& $4421\pm 4$  & 4499.4  \\\hline
5& \;$\psi(4660)\;$& $4633 \pm 7 $     & 4806.3    \\\hline
6*& \;$\psi(6S)\;$&  ------   & \clt{4889.5${}^\star$} \\\hline
7*& \;$\psi(7S)\;$&  ------    &  \clt{5111.4${}^\star$}  \\\hline
8*& \;$\psi(8S)\;$& ------    & \clt{5318.7${}^\star$}  \\\hline
\hline\hline
\end{tabular}
\caption{Table \ref{scalarmasses2} completed with the higher $n$  resonances (with an asterisk) of the $\psi$ meson family. The extrapolated masses for $n=6, 7, 8$ in the fourth column indicated with a `` ${}^\star$ '' specify that they are result of the concomitant use of DCE Regge  trajectories (\ref{itp2}, \ref{itq12}), interpolating the experimental masses of $\psi$ mesons from $n=1,\ldots,5$.} \label{scalarmasses20}
\end{center}
\end{table}
\noindent In Table \ref{scalarmasses20} the  $\psi(4660)$ ($n=5$) resonance has a  theoretically-predicted mass that is slightly higher than the mass obtained for the $\psi(6S)$ ($n=6$) resonance. The point here is that up to the $\psi(4660)$ state, the masses in the fourth column of Table \ref{scalarmasses20} were derived using exclusively AdS/QCD, whereas the masses of the $\psi(6S)$, $\psi(7S)$, and $\psi(8S)$ resonances were here derived using interpolation of experimental values of the masses, namely, the third column of  Table \ref{scalarmasses20}. Hence the monotonic profile of the masses is satisfied, since one must read in Table \ref{scalarmasses20} the experimental masses of the $J/\psi(1S) $, $\psi(2S)$, $\psi(4040)$, $\psi(4415)$, $\psi(4660)$ states and, subsequently, the masses of the $\psi(6S)$, $\psi(7S)$, and $\psi(8S)$ resonances in the fourth column.

The $\omega$ isovector meson family can be analyzed, using the same approach. The  $\omega(7S)$ meson resonance can have its mass derived by considering the excitation level  $n=7$ in Eq. (\ref{itp3}),  yielding the DCE equals \clt{21.478 nat}. Next, replacing it in the left hand side of Eq. (\ref{itq13}), and posteriorly solving for the $m$ variable implies that the  $\omega(7S)$ resonance mass equals \clt{$m_{\omega(7S)}= 2499.7$ MeV}. 
Therefore accomplishing the same approach to the $n=8$ excitation level yields the DCE of the $\omega(8S)$ wave resonance to be equal to \clt{24.087 nat}, when employing the DCE Regge  trajectory (\ref{itp3}). Subsequently resolving Eq. (\ref{itq13})  \clt{$m_{\omega(8S)}= 2649.3$ MeV}. 
Similarly,  the $n=9$ excitation level can now be used in Eq. (\ref{itp1}). The DCE of the $\omega(9S)$ resonance equals \clt{26.960 nat}, using the Regge  trajectory (\ref{itp3}). Then solving Eq. (\ref{itq13}) for this value, the mass of the  $\omega(9S)$ meson resonance  reads \clt{$m_{\omega(9S)}= 2789.5$ MeV}. 
These outcomes  are shown in Table 
\ref{scalarmasses30}. 

\begin{table}[H]
\begin{center}\begin{tabular}{||c|c||c|c||}
\hline\hline
$n$ & State & $M_{\scalebox{.67}{\textsc{Experimental}}}$ (MeV)  & $M_{\scalebox{.67}{\textsc{Theory}}}$ (MeV) \\
       \hline\hline
\hline
1 &\;$\omega(782)\;$ & $782.65\pm0.12$ & 981.43   \\ \hline
2 &\;$\omega(1420)\;$ & $1410 \pm 60$ & 1374  \\ \hline
3& \;$\omega(1650) \;$& $1670\pm 30$       & 1674     \\\hline
4& \;$\omega(1960) \;$& $1960\pm25$  & 1967  \\\hline
5& \;$\omega(2205) \;$& $2205 \pm 30 $     & 2149   \\\hline
6& \;$\omega(2290)\;$&  $2290 \pm 20$   & 2499.7 \\\hline
7*& \;$\omega(7S)\;$& ------     &  \clt{2358.4${}^\star$}  \\\hline
8*& \;$\omega(8S)\;$& ------      & \clt{2649.3${}^\star$}   \\\hline
9*& \;$\omega(9S)\;$& ------     & \clt{2789.5${}^\star$}  \\\hline
\hline\hline
\end{tabular}
\caption{Table \ref{scalarmasses3} completed with the higher $n$  resonances (with an asterisk) of the $\omega$ meson family. The extrapolated masses for $n=7, 8, 9$ in the fourth column indicated with a `` ${}^\star$ '' specify that they are result of the concomitant use of DCE Regge  trajectories (\ref{itp3}, \ref{itq13}), interpolating the experimental masses of $\omega$ mesons from $n=1,\ldots,6$.} \label{scalarmasses30}
\end{center}
\end{table}
\noindent The state $\omega(2330)$ with  $I^G=0^-$ and $J^{PC}=1^{--}$ quantum numbers, reported in PDG 2020 \cite{pdg}, has mass  $2330\pm30$ MeV and may be immediately identified to the 
$\omega(7S)$ isovector meson resonance.  \clt{Also, the state $X(2632)$ with mass  $2635.2$ MeV might be matched with the 
$\omega(8S)$ isovector meson resonance. }

Finally, the $\phi$ isovector meson family can be studied by this method. The  $\phi(4S)$ resonance  corresponds to the $n=4$ excitation level. 
When putting  $n=4$ back into Eq. (\ref{itp4}), the value   \clt{19.870 nat} of the DCE is derived. Hence, replacing it into Eq. (\ref{itq14}), the  $\phi(4S)$ resonance mass reads \clt{$m_{\phi(4S)}= 2560.6$} MeV.  For $n=5$, the DCE of the $\phi(5S)$ resonance is \clt{22.441 nat}, when the DCE Regge  trajectory (\ref{itp4}) is employed. Therefore  Eq. (\ref{itq14}) can be solved for it, yielding the  $\phi(5S)$ resonance mass \clt{$m_{\phi(5S)}= 2913.3$ MeV}. 
Analogously,  the $n=6$ excitation level in Eq. (\ref{itp4}) yields the DCE of the $\phi(6S)$ resonance equals \clt{25.230} nat. Subsequently resolving Eq. (\ref{itq14}) for it, the $\phi(6S)$ resonance mass \clt{$m_{\phi(6S)}=3232.4$ MeV} is acquired.

\begin{table}[H]
\begin{center}
\begin{tabular}{||c|c||c|c||}
\hline\hline
$n$ & State & $M_{\scalebox{.67}{\textsc{Experimental}}}$ (MeV)  & $M_{\scalebox{.67}{\textsc{Theory}}}$ (MeV) \\
       \hline\hline
\hline
1 &\;$\phi(1020)\;$ & $1019.461\pm0.016$ & 1139.43   \\ \hline
2 &\;$\phi(1680)\;$ & $1680 \pm 20$ & 1583  \\ \hline
%3& \;$\phi(1650) \;$& $1670\pm 30$       & 1921     \\\hline
3& \;$\phi(2170) \;$& $2160\pm80$  & 2204  \\\hline
4*& \;$\phi(4S) \;$& ------    &\clt{2560.6${}^\star$}  \\\hline
5*& \;$\phi(5S)\;$&  ------  & \clt{2913.3${}^\star$} \\\hline
6*& \;$\phi(6S)\;$&  ------    & \clt{3232.4${}^\star$}  \\\hline
%8& \;$\eta(2225)\;$& $2221^{+13}_{-10}$      & 2193.6   \\\hline
%9& \;$\eta(2320)\;$& $2320\pm 15$      & 2289.4  \\\hline
\hline\hline
\end{tabular}
\caption{Table \ref{scalarmasses4} completed with the higher $n$  resonances (with an asterisk) of the $\omega$ meson family. The extrapolated masses for $n=4, 5, 6$ in the fourth column indicated with a `` ${}^\star$ '' specify that they are result of the concomitant use of DCE Regge  trajectories (\ref{itp4}, \ref{itq14}), interpolating the experimental masses of $\phi$ mesons from $n=1,\ldots, 3$.} \label{scalarmasses40}
\end{center}
\end{table}
\noindent  \clt{The $\phi(6S)$ isovector meson state, whose mass is  predicted to be 2913.3 MeV, might correspond to the  $X(2350)$ element with experimental mass  $3250\pm 28$ MeV.}

\section{Concluding remarks and perspectives}\label{iv}

 DCE Regge trajectories represent a successful approach to mass spectroscopy, in approaches of  AdS/QCD. Here the four isovector meson families were scrutinized in this setup. Their underlying information entropy content played an important role in deriving the mass spectra of the next generation of isovector meson resonances, with a higher excitation level. To accomplish it, a shift of the quadratic dilaton (\ref{devia}) was used. One of the advantages of using  DCE-based techniques is that the computational apparatus is simple when compared to other phenomenological approaches of AdS/QCD, including the lattice one. Another purpose of the DCE approach for isovector mesons is the interpolation of their experimental mass spectra of already detected states in PDG \cite{pdg}, to derive the mass spectra of the next generation of isovector meson resonances. This is implemented by the concomitant analysis of the DCE Regge trajectories Eqs. (\ref{itp1}) -- (\ref{itp4}) and (\ref{itq11}) -- (\ref{itq14}), pairwise respectively to the $\Upupsilon$, the $\psi$, the $\omega$, and the $\phi$ isovector meson families.  
As the DCE is equivalent to the information chaoticity carried by a message, one can additionally interpret the isovector meson families as byproducts of experimental multiparticle production processes. In high-energy collisions encompassing quarks and gluons, chaos in QCD gauge dynamics may set in, and only
 particles in final states can be measured. The chaotic profile of collisions that produce isovector meson excitations can be then 
quantified by the loss of information at the end of the collision processes. This quantitative aspect was here shown to be encompassed by the DCE, providing experimental determination by the derived mass spectra of isovector meson states with higher excitation numbers.

 Another important feature of this method is the configurational stability of the isovector meson resonances. Tables \ref{scalarmasses5} -- \ref{scalarmasses8} respectively display the DCE of the $\Upupsilon$, the $\psi$, the $\omega$, and the $\phi$ isovector meson families.  Besides,  from the paragraph that precedes Table \ref{scalarmasses20} up to the end of Sec. \ref{sec2}, the values of the DCE of higher excitation level resonances are shown monotonically to increase as a function of their excitation level. This implies a configurational instability for higher excitation level resonances, irrespectively of the isovector meson family. It also corroborates the fact that lower excitation level resonances have been already detected in experiments \cite{pdg}, as they are more dominant for the higher excitation level ones. Equivalently, given any fixed family among the $\Upupsilon$, the $\psi$, the $\omega$, and the $\phi$ isovector ones, the less massive the  meson state, the less unstable they are, under the prism of  DCE tools. The extrapolation of the mass spectra of isovector meson higher excitations can be formally implemented for an arbitrary excitation number $n$ with good accuracy. However, Tables \ref{scalarmasses10} -- \ref{scalarmasses40} consider three excitation numbers beyond the experimental data, in each one of the four isovector meson families. Beyond that, further states  are not explored, since they are configurationally very unstable, with high values of DCE. As such, they are unlikely to be experimentally detected, at least with proposed experiments that have been currently run.

AdS/QCD-based models that report quarkonia states were  investigated  in Ref. \cite{Braga:2015jca}, whose DCE was computed in Ref. \cite{Braga:2017fsb}, using the standard quadratic soft wall dilaton. This has been also emulated to the finite temperature and density scenario  \cite{Braga:2016wkm,Braga:2017oqw}. In heavy-ions collisions that  produce the quark-gluon plasma (QGP),  quarkonia are originated also percolating  into the QGP and, therefore, dissociating as an effect of the
hot temperature. The next step of our studies consists of extending this model with the shifted dilaton potential \eqref{devia}. In this way, some thermal features of  mesons, whose constituents are heavy quarks, can be investigated, for example, when dissociated in  QGP.
Also,  decay constants can be studied in finite temperature models. As a preliminary study of the DCE underlying the QGP was already studied in Ref. \cite{daSilva:2017jay}, one can also investigate the $\Upupsilon$ mesons suppression in heavy-ion
collisions. As the $\Upupsilon$ mesons binding energy is  large when compared to the one of $\psi$ resonances, the dissociation energy of $\Upupsilon$ mesons in the QGP is also expected to be higher \cite{Benzahra:1999iv}. The absorption of $\Upupsilon$ and $\psi$ resonances  in the hadronic matter then also consists of relevant directions of future investigations. 
\medbreak
\paragraph*{Acknowledgments:}RdR is grateful to FAPESP (Grants No. 2017/18897-8 and No. 2021/01089-1) and the National Council for Scientific and Technological Development -- CNPq (Grants No. 303390/2019-0 and No. 406134/2018-9), for partial financial support. 

\end{document}